\documentclass[12pt]{elsart}
\usepackage{graphicx}
\usepackage {amssymb}
\begin{document}
\begin{frontmatter}
\title{Anomalies of conductivity behavior near the
paramagnetic-antiferromagnetic transition in single-crystals
La$_{2}$CuO$_{4+\delta}$}

\author{B. I. Belevtsev\corauthref{cor}},
\author{N. V. Dalakova},
\author{A. S. Panfilov},
\author{E. Yu. Beliayev}

\address {B. Verkin Institute for Low Temperature Physics and
Engineering, National Academy of Sciences, Kharkov, 61103,
Ukraine}

\corauth[cor]{Corresponding author. Email address:
belevtsev@ilt.kharkov.ua. Fax: ++380-572-335593. Phone:
++380-57-3410963.}

\begin{abstract}
The temperature dependences of resistance, $R(T)$, of two
single-crystals La$_{2}$CuO$_{4+\delta}$ samples have been studied
with the aim to detect a possible change in the $R(T)$ behavior
induced by paramagnetic-antiferromagnetic (PM-AFM) transition. One
of the samples with $\delta \lesssim 0.01$, was fairly homogeneous
in oxygen distribution (not phase-separated) with N\'{e}el
temperature $T_{N}\approx 266$~K. Conductivity of this sample has
been determined by Mott's variable-range hopping below $T_N$. The
other, far less resistive, sample with $\delta \approx 0.05$, was
inhomogeneous (phase-separated) showing both PM-AFM ($T_N\approx
205$~K) and superconducting ($T_c\approx 25$~K) transitions. It is
found that for the homogeneous sample  the resistivity decreases
above $T_N$ far faster with temperature than below it (for both
directions of measuring current, parallel and perpendicular to
basal CuO$_2$ planes). A similar behavior of conductivity near
PM-AFM transition is also found for the phase-separated and less
resistive sample. In this case a clear kink in $R(T)$ curve near
$T_N\approx 205$~K can be seen. Furthermore, a transition to
metallic ($dR/dT>0$) behavior occurs far enough above $T_N$.  The
observed behavior of the samples studied is related to increased
delocalization of charge carriers above $T_N$. This is in
accordance with decrease in the AFM correlation length and
corresponding enhancement of the hole mobility above $T_N$ known
for low-doped lanthanum cuprates.

\end{abstract}

\end{frontmatter}

\section{Introduction}
Research on cuprate high-$T_{c}$ superconductors  is one of the
topical problems of physics of solids. The nature of
superconductivity in these compounds (discovered more than 20
years ago) is still not entirely clear. It is however evident that
their magnetic and superconducting properties  are closely
related. In the normal state, the conducting properties can also
be considerably dependent on the magnetic state of the system. For
example, in single crystal La$_{2}$CuO$_{4}$ the metamagnetic
antiferromagnetic (AFM) - weak ferromagnetic transition causes a
sharp increase in the conductivity \cite{tineke,boris}.
Investigations of the interaction between the charge carriers and
the magnetic subsystem are important not only for high-$T_{c}$
superconductors but generally for a wide range of other magnetic
conductors and semiconductors as well.
\par
The above consideration has stimulated this study of transport and
magnetic properties of single-crystal La$_{2}$CuO$_{4+\delta}$
cuprates near the N\'{e}el temperature $T_{N}$. The objective was
to detect possible correlations of these properties. Non-doped
La$_{2}$CuO$_{4}$ is an AFM insulator ($T_{N} \approx $ 320 K)
\cite{kastner}. Its doping with excess oxygen ($\delta \ne 0$)
gives rise to charge carriers (holes) and supresses the AFM order
(decreases $T_{N}$). The properties of a single crystal
La$_{2}$CuO$_{4}$ are to a great extent determined by its crystal
and magnetic structures \cite{kastner,vakh}. In the temperature
region of our interest (below 430 K), La$_{2}$CuO$_{4}$ has an
orthorhombic perovskite-type lattice consisting of alternating
CuO$_{2}$ and La$_2$O$_2$ layers. In the $Bmab$ space group the
CuO$_{2}$ layers are perpendicular to the $c$-axis and parallel to
the $ab$-plane \cite{kastner}. The charge carriers (holes) are
believed \cite{kremer} to be mainly of oxygen character. The
excess oxygen (when $\delta \ne 0$) is located in the
La$_{2}$O$_{2+\delta}$ layers between the adjacent CuO$_{2}$
planes \cite{kremer}. This provides hole delocalization from the
CuO$_{2}$ planes and thus leads to the three-dimensional (3D)
character of conductivity. The magnetic structure is formed by the
copper ions $d^{9}$Cu$^{2+}$ with spin $S=0.5$. The spins at the
neighboring Cu$^{2+}$ sites are oppositely directed and aligned
along the orthorhombic $b$-axis (Cu-Cu direction) in the
$bc$-plane. The spins are canted ($\approx 0.17^{\circ}$) with
respect to the $b$-axis. Each CuO$_{2}$ layer has a weak
ferromagnetic moment perpendicular to the layer. Below $T_N$, the
directions of the moments in neighboring CuO$_{2}$ planes are
opposite, so that the system as a whole is a 3D AFM \cite{tineke}.
\par
It is well known that in metallic conductors the transition from
paramagnetic (PM) into AFM state usually implies a decrease in
magnetic part of electric resistance that causes a kink in
temperature dependence of resistance $R(T)$ \cite{gratz} or even
an appreciable drop of the resistance on transition to the AFM
state \cite{boris2}. In magnetic semiconductors, the influence of
spin ordering at $T\leq T_N$ on conductivity is still not
understood sufficiently well. It is, however, known
\cite{metf,nagaev} that conductivity of these compounds
demonstrates often a weak kink in the $R(T)$ at $T\approx T_N$. It
can be expected that similar effect takes place in conductivity of
La$_{2}$CuO$_{4+\delta}$ as well.
\par
The effect of PM-AFM transition  on temperature dependence of
resistance has been noted in studies of some underdoped
high-$T_{c}$ cuprates. For example, for single-crystal
YBa$_2$Cu$_3$O$_{6+x}$ a clear feature near $T_N$ has been found
in the out-of-plane resistivity $\rho_{c}(T)$ (that is when
current is parallel to the $c$-axis), whereas, the in-plane
$\rho_{ab}(T)$ (measured with current parallel to CuO$_2$ layers)
shows no feature at $T=T_N$ \cite{lavrov}. The same behavior was
found for lightly doped La$_{2-x}$Sr$_x$CuO$_4$ \cite{komi}. In
both cases this reflects strong anisotropy of
quasi-two-dimensional (2D) conductivity of these layered
compounds.
\par
The behavior of conductivity in non- or low-doped
La$_{2}$CuO$_{4+\delta}$ near $T_{N}$ was approached previously
\cite{hund,parf}. In Ref. \cite{hund} the temperature dependence
of electric resistance $R(T)$ shows an anomaly near or not very
far from $T_{N}$ in the samples with $T_{N}$ in the range 250--310
K, but these anomalies appeared to be determined by the phase
separation \cite{ryder,gren,rada,chou,birg,tranq} upon cooling
into oxygen-poor ($\delta\approx 0$) and oxygen-rich ($\delta
> 0$) phases. In study of Ref. \cite{parf} anomalous
behavior of resistance was searched for at $T \approx T_{N}$ by
examining the temperature behavior of the activation energy of
hopping conduction. It was found clearly in a sample with $T_{N}
\approx 300$~K, but for other samples studied (with $T_N$ equal to
318 K and 275~K) it is hardly can be spoken about some definite
resistance anomaly near $T_N$. The previous investigations thus,
on the one hand, suggest a possibility of anomalous $R(T)$
behavior near $T_{N}$ in La$_{2}$CuO$_{4+\delta}$ crystals
low-doped with excess oxygen, but, on the other hand, some crucial
questions concerning this effect remained unanswered. For example,
it is important to know how universal the anomaly is and what
factors are responsible for it. We believe that the results of
this study can be helpful in answering these questions.

\section{Samples and experimental details}  Two samples of
single crystals La$_{2}$CuO$_{4+\delta}$ (both grown at the
Institute of Solid State and Semiconductor Physics, Minsk,
Belorussia by S. N. Barilo group) were used in this study. First
of them (No. 1) was with dimensions about 4$ \times $4$ \times
$3.3 mm$^{3}$. The temperature of AFM transition ($T_{N} = 266$~K)
was determined from the position of the peak in the temperature
dependence of dc magnetic susceptibility $\chi$(T) (Fig. 1) in the
magnetic field directed along the $c$-axis. The measurements were
made in a Faraday-type magnetometer. According to the known phase
diagram for La$_{2}$CuO$_{4+\delta}$ \cite{gren,chou,birg,tranq}
this value of $T_N$ corresponds to $\delta \lesssim 0.01$. At this
low content of excess oxygen the La$_{2}$CuO$_{4+\delta}$ sample
is expected to be in a rather homogeneous non-metallic phase state
(no mixed-state or substantial phase separation effect)
\cite{gren,chou,birg,tranq}.
\par
The second sample (No. 2) (with the size about 1.5$ \times $2$
\times $2 mm$^{3}$) was studied previously in Ref. \cite{gnezd}
where $\delta \simeq 0.05$ was estimated. The sample was shown
\cite{gnezd} to be inhomogeneous, consisted of AFM and
superconducting phases. Temperature dependence of the dc magnetic
susceptibility $\chi(T)$ (Fig. 2), measured in this study, shows
both the AFM ($T_N=205\pm 2$~K) and superconducting ($T_c\approx
25$~K) transitions, reflecting its phase separated state. The
values of $T_N$ and $T_c$ correspond well to those measured in
Ref. \cite{gnezd}.
\par
The temperature dependences of resistivity, $\rho (T)$, were
measured with a direct measuring current using the four-probe
method in zero magnetic field ($H=0$). To produce contacts for the
current and potential leads, silver was deposited onto the contact
pads and then thin gold wires were glued with silver paste.
Current was passed both along and across the CuO$_{2}$ planes. The
measuring current was selected (using the measured current-voltage
characteristics) so that the Ohm's law could be obeyed in the
whole $T$-range of the measurement. The measuring current $J$ for
sample No. 1 ($T_N\approx 266$~K) was varied within 3--10 $\mu$A.
The less resistive sample No. 2 ($T_N\approx 205$~K) was measured
at higher current $J\leq 100$~$\mu$A.

\section{Results and discussion}
\subsection{Sample La$_{2}$CuO$_{4+\delta}$ with $\delta \lesssim 0.01$
and $T_N\approx 266$~K}

The obtained dependences $\rho (T)$ (presented as $\log \rho$ {\it
vs.} $T^{-1/4}$) for the in-plane (current $J$ parallel to the
CuO$_2$ planes) and out-of-plane ($J\|c$) transport are shown in
Fig. 3. As expected for layered cuprates, the out-of-plane
resistivity is far higher than the in-plane one. Both curves have
a clear kink (or turn) near $T_{N}\approx 266$~K. Below $T_N$, the
curves follow the Mott's law for variable-range hopping

\begin{equation}
\label{eq1} \rho \approx \rho_{0} \exp\left( {\frac{{T_{0}}
}{{T}}} \right)^{1/4},
\end{equation}

\noindent which agrees with the previous data for
La$_{2}$CuO$_{4+\delta}$ \cite{boris,kast2,boris3}. The value of
exponent (1/4) corresponds with that expected for a 3D system
\cite{mott}. Below $T_{N}$, the characteristic temperatures
$T_{0}$ [see Eq.(\ref{eq1})] for the curves $\rho (T)$ in Fig. 3
are $3.3\times 10^{5}$~K and $7.7\times 10^{4 }$~K for the
in-plane and out-of-plane currents, respectively. The localization
length $L_{c}$ can be estimated from the expression $kT_{0}
\approx 16/[N(E_{F})L_{c}^{3}]$ \cite{mott}, where $N(E_{F})$ is
the density of charge-carrier states at the Fermi level. Using
$N(E_{F}) = 2.8 \times 10^{46}$~J$^{-1}$m$^{-3}$ \cite{jarl}, the
values $L_{c} \approx 0.5$~nm and $L_{c}\approx 0.84$~nm can be
obtained for the in-plane and out-of-plane directions of the
current, respectively, for $T \leq 266$~K. This corresponds to
previous estimates of the value of $L_c$ in La$_{2}$CuO$_{4}$
below 200~K \cite{kast2,boris3}. Above $T_N$, the investigated
$T$-interval is not wide enough (on the inverse temperature scale)
to determine precisely functional dependence $\rho (T)$ of an
exponential type like Eq. (\ref{eq1}). If, nevertheless, to apply
dependence (\ref{eq1}) formally for comparison with $\rho (T)$
above $T_N$, the calculated values of $L_c$ turn out to be too
small (less than interatomic O-O or Cu-Cu distances) to believe in
variable-range hopping in this temperature range. At the same time
the simple exponential dependence $\rho (T) \propto
\exp(E_{a}/kT)$ is obviously inconsistent with the experimental
results in the whole temperature interval studied.
\par
In Fig. 4 temperature behavior of resistivity is presented as
$\log \rho$ {\it vs.} $T$ that shows more evidently the change in
$\rho(T)$ behavior near $T_N$. It is seen that below $T_N$ the
$\rho(T)$ follows the Mott's law (as indicated above) but above
$T_N$ the resistivity decreases with temperature far faster than
below it. This is in contrast with $\rho(T)$ behavior near $T_N$
in metallic conductors where resistivity increases when going from
AFM to PM state \cite{gratz,boris2,metf}. At the same time, in
some magnetic semiconductors a kink in $\rho(T)$ near $T_N$
similar to that shown in Fig.~4 is observed \cite{nagaev}. An
increasing in activation energy of conducting electrons at AFM-PM
transition was proposed as a possible explanation but it was not
considered as a general rule \cite{nagaev}. For the sample studied
no some definite activation energy exists below or above $T_N$, so
we cannot apply this explanation.
\par
Magnetoresistance (MR) was measured in this study using a rotating
Kapitza electromagnet with the highest field 1.7~T. In the
interval 170--400 K in fields up to 1.7~T, the MR was very low
($|\Delta R(H)/R(0)|< 10^{-3}$). This does not permit to judge
about an anomaly in the MR behavior near $T_{N}$. It can not be
excluded, however, that in rather strong fields MR of
La$_{2}$CuO$_{4+\delta}$ could reveal some anomalous behavior near
$T_{N}$. The possibility of such anomaly caused by the spin
disorder influencing the mobility of charge carriers was discussed
in \cite{belov}. An anomaly in the out-of plane MR ($J\parallel
c$) upon crossing $T_N$ with changing in temperature was found
actually in single-crystal YBa$_2$Cu$_3$O$_{6+x}$  for both the
in-plane and out-of-plane orientations of the field \cite{lavrov}.
A rather high magnetic field, $H=16$~T, was needed, however, in
this case to increase MR magnitude up to about $1\%$ below $T_N$
and to reveal clearly enough this effect \cite{lavrov}.
\par
The effect of magnetic ordering on the hopping conduction of
magnetic semiconductors should be analyzed taking into account the
type of magnetic order and the origin of charge carriers. The
nature of holes in La$_{2}$CuO$_{4+\delta}$ is not completely
clear so far (see the Discussion in \cite{boris}). Although it is
commonly supposed that the holes in La$_{2}$CuO$_{4+\delta} $ are
mainly of oxygen character \cite{kremer}, the overlapping of the
$d$- and $p$-orbitals and the $d$- and $p$-band hybridization
\cite{hott} can also have an appreciable effect on the nature and
motion of holes. It is known that in high-$T_c$ cuprates the AFM
order and the charge carriers (holes) are antagonistic to each
other \cite{keimer}. On the one hand, at $T < T_{N}$ holes of any
type, in addition to lattice distortion and the related polaron
effects, can cause considerable distortion of the AFM order in
cuprates \cite{carl}. The hole can have a spin and its motion can
add to the disturbance of the AFM order (frustration effect)
\cite{carl}. On the other hand, the AFM order impedes the motion
of the holes.
\par
Above $T_{N}$, the long-range 3D AFM order starts to decay quickly
\cite{keimer} though the 2D AFM correlations can persist in the
CuO$_{2}$ planes even far above $T_{N}$ \cite{kastner}. It can be
suggested, therefore, that in cuprates with weak coupling between
the CuO$_{2}$ layers (quasi-2D conductors with strong conductivity
anisotropy) some feature at $T\simeq T_N$ is expected only for the
out-of-plane conductivity, when  the interlayer exchange coupling
is destroyed above $T_N$. For the in-plane conductivity nothing
special is to expect since 2D intralayer AFM correlation still
exists even far above $T_N$. This type of behavior was really seen
in quasi-2D cuprates like YBa$_2$Cu$_3$O$_{6+x}$ \cite{lavrov} and
lightly-doped La$_{2-x}$Sr$_x$CuO$_4$ \cite{komi}, where features
at $T\approx T_N$ were found only in the out-of-plane resistivity
but not in the in-plane one. By contrast, the observed variations
in conductivity near $T_N$ in the sample studied manifest that the
destruction of the 3D AFM order by thermal fluctuations impacts on
conductivity in both the in-plane and out-of-plane directions
(Figs. 3 and 4). This is not surprising due to specific influence
of oxygen holes which leads to 3D character of hopping
conductivity in La$_{2}$CuO$_{4+\delta}$ with excess oxygen
\cite{boris,kremer} as it is mentioned above.
\par
In low-doped La$_{2-x}$Sr$_{x}$CuO$_{4}$ the hole mobility
increases as the AFM correlation length $\xi _{AF}$ decreases,
i.e. as the AFM order grows weaker \cite{ando}. At the same time,
it is known \cite{kastner,keimer} that in La$_{2}$CuO$_{4+\delta}$
and low-doped La$_{2-x}$Sr$_{x}$CuO$_{4}$ the length $\xi _{AF}$
is practically independent on temperature below $T\approx $ 300 K.
However, above $T_N$, it decreases drastically with rising
temperature \cite{kastner,keimer}. Thus, the change in the
behavior of conductivity at temperature exceeding $T_{N}$ (Figs. 3
and 4) can be related to the decrease in the AFM correlation
length and the corresponding enhancement of the hole mobility in
this temperature region.
\par
It should be noted that some of the authors of this article have
observed more than once the $R(T)$ of La$_{2}$CuO$_{4+\delta}$
(with $T_{N}$ in the range 180--280~K) behaving near
\textit{T}$_{N}$ as in Fig. 3 (e.g., see \cite{boris}). The study
of Ref. \cite{boris} was done, however, only up to 300 K, so that
this effect does not appear quite clearly. In this study the
temperature interval was extended to 430 K. This made possible to
see the change in behavior of $R(T)$ near $T_N$ more clearly and
allowed us to relate it to the decrease in the AFM correlation
length above $T_{N}$.
\par
It is significant as well that behavior of $R(T)$ near N\'{e}el
temperature quite similar to that shown in Fig.~3 was found in
other magnetic oxides like manganites
(La$_{0.25}$Ca$_{0.75}$MnO$_3$ \cite{zheng}) or cuprates with
variable-range hopping transport
(Y$_{0.37}$Pr$_{0.63}$Ba$_2$Cu$_3$O$_7$ \cite{jiang},
YBa$_2$Cu$_3$O$_{6+x}$ \cite{lavrov}, La$_{2-x}$Sr$_x$CuO$_4$
\cite{komi}). It seems therefore that particular features of the
change in temperature behavior of hopping conduction of
La$_{2}$CuO$_{4}$ near $T_N$ found in this study (variable-range
hopping below $T_N$ and progressive decreasing in resistivity with
temperature above $T_N$) is perhaps a common property of magnetic
semiconducting oxides with perovskite-like lattice.

\subsection{Sample La$_{2}$CuO$_{4+\delta}$ with $\delta \approx 0.05$
and $T_N\approx 205$~K}

As it is mentioned above, this sample is inhomogeneous consisting
of a mixture of AFM and superconducting phases, which is reflected
in the temperature behavior of the dc susceptibility $\chi (T)$
(Fig. 2). Let us consider at first the temperature dependence
$\rho (T)$ recorded in the in-plane current direction (Fig. 5)
which is found to be in complete concordance with $\chi (T)$,
showing resistive superconducting transition with $T_c\approx
25.5$~K and clear feature (kink) near $T_N\approx 205$~K.
\par
On the whole, this sample (No. 2) is much ($\approx 10^3$) less
resistive than the sample No.~1 with $T_N\approx 266$~K (compare
Figs. 3 and 5). Nevertheless, the variable-range hopping [$\ln
\rho(T)\propto T^{-1/4}$] is also revealed in this sample  in the
range between $T_c$ and $T_N$ (Fig. 5). In the same way as for the
sample No. 1, an increased drop in resistivity with temperature is
seen above $T_N$; furthermore, far enough above $T_N$ (at
$T_{min}\approx 350$~K, where resistance minimum shows up) a
transition from non-metallic ($d\rho/dT<0$) to metallic
($d\rho/dT>0$) temperature behavior of $\rho(T)$ takes place.
\par
Resistive superconducting transition in the in-plane direction is
fairly sharp (Fig.~5) with $T_c\approx 25.5$~K defined as the
temperature of the inflection point of $\rho(T)$ curve (which was
found using derivative). In low-temperature range of the resistive
transition (at $T\leq 25$~K) some clear ``shoulder'' in $\rho(T)$
can be seen, as that commonly observed in granular (low- and
high-$T_c$) superconductors with a rather weak connection between
grains \cite{usp,boris4}. Thus it can be thought that
superconducting regions in this inhomogeneous sample make some
sufficiently continuous chain to show the rather sharp resistive
superconducting transition. On the other hand, however, some links
of this chain are weak which determines the shoulder in
low-temperature part of the resistive transition curve
\cite{usp,boris4}. The indicated features of the resistive
transition (the ''shoulder`` together with non-zero resistance
even at lowest temperature of this study) reflect unambiguously
two-phase (phase-separated) state of the sample, in which some of
the superconducting ''grains`` (clusters)  are isolated within AFM
matrix.
\par
The resistivity in the out-of-plane $(J\|c)$ direction is (same as
for sample No.~1) much larger than that for the in-plane one (Fig.
6). Nevertheless, the temperature dependence of resistivity has
main features similar to those in the in-plane current direction.
Superconductivity shows itself as a considerable drop of
resistance when approaching  the temperature $T_c\approx 25.5$~K
from above. The resistance, however, does not come down to zero,
which is an evidence of higher phase inhomogeneity in this
direction, so that a continuous superconducting chain is not
formed. Change in $\rho(T)$ behavior after elevation of
temperature above $T_N$ is identical to that for the in-plane
current direction. Even a transition from nonmetallic to metallic
$\rho(T)$ behavior takes place (resistance minimum at
$T_{min}\approx 385$~K).
\par
\subsection{Summary}
The results obtained demonstrate clear interrelation between the
magnetic and conducting properties of low-doped
La$_{2}$CuO$_{4+\delta}$. Both samples, although significantly
different in resistivity and $T_N$, show the same characteristic
changing in transport properties above $T_N$: system becomes less
resistive and more metallic. In sample No. 2 even a nonmetal-metal
transition takes place far enough above $T_N$. This conforms with
known theoretical concepts and some experiments
\cite{keimer,carl,lavrov2}. According to these, the AFM order
enhances hole localization while thermal destruction of the AFM
order induces delocalization of charge carriers,  so that above
$T_N$ a system can approach metallic state with increasing
temperature. At $T\approx T_N$ only interplane AFM order
disappears; whereas, the in-plane AFM can survive even rather far
above $T_N$. At the same time, the AFM correlation length $\xi
_{AF}$ decreases drastically with temperature above $T_N$
\cite{kastner,keimer} and this should be accompanied with
enhancement of the hole mobility \cite{ando}. Strong increase in
conductivity above $T_N$ can lead in some cases to nonmetal-metal
transition as it is observed in this study. These effects are seen
in both the in-plane and out-of-plane directions which is
determined by 3D character of hopping conduction in
La$_{2}$CuO$_{4+\delta}$.

\ack The authors are indebted to V. N. Savitsky and V. P.
Gnezdilov for providing with the La$_{2}$CuO$_{4+\delta}$ samples
for this study.

\newpage

\centerline{\bf{Figures}} \vspace{12pt}

Figure 1. Temperature dependence of the dc magnetic susceptibility
$\chi$ in the field $H=0.543$~T parallel to the crystallographic
axis $\mathbf{c}$ of single crystal La$_{2}$CuO$_{4+\delta}$
(sample No. 1, $\delta \lesssim 0.01$). \vspace{15pt}

Figure 2. Temperature dependence of the dc magnetic susceptibility
$\chi$ in the field $H=0.83$~T parallel ($\parallel$) and
perpendicular ($\perp$) to the crystallographic axis $\mathbf{c}$
of single crystal La$_{2}$CuO$_{4+\delta}$ (sample No. 2, $\delta
\approx 0.05$). The inset shows diamagnetic response below the
superconducting transition temperature $T_c\approx 25$~K.
\vspace{15pt}

Figure 3. Temperature dependence of resistivity (presented as
$\log \rho$ {\it vs.} $T^{-1/4}$) of single crystal
La$_{2}$CuO$_{4+\delta}$ (sample No. 1, $T_{N} = 266$~K) for the
in-plane (current $J$ parallel to the CuO$_2$ planes) and
out-of-plane ($J\|c$) directions of the measuring current. In both
cases, solid lines present Mott's law (Eq. 1) which is obeyed
below $T_N$. \vspace{15pt}

Figure 4. (Color online) Temperature dependence of resistivity of
sample No.~1 (presented as $\log \rho$ {\it vs.} $T$) for
measuring current parallel to the $c$-axis. The dashed line
presents Mott's law (Eq. 1). It is seen that above $T_N$
progressive deviation from this law sets in. \vspace{15pt}

Figure 5. Temperature dependence of resistivity of single crystal
La$_{2}$CuO$_{4+\delta}$ (sample No. 2, $\delta \approx 0.05$,
$T_{N} \approx 205$~K) for the in-plane (current $J$ parallel to
the CuO$_2$ planes) direction of the measuring current. The inset
presents this dependence as $\ln \rho$ {\it vs} $T^{-1/4}$ to show
hopping conductivity in the range between $T_c$ and $T_N$. The
value of $T_c\approx 25.5$~K is defined from  the position of a
peak in the temperature behaviour of the derivative $d\rho/dT$ in
the region of superconducting transition.\vspace{15pt}

Figure 6. Temperature dependence of resistivity of single crystal
La$_{2}$CuO$_{4+\delta}$ (sample No. 2, $\delta \approx 0.05$,
$T_{N} \approx 205$~K) for the out-of-plane ($J\|c$) direction of
the measuring current. The inset presents this dependence as $\ln
\rho$ {\it vs} $T^{-1/4}$. Positions of $T_N$ and $T_c$ are shown
by arrows.

\newpage
\begin{figure}
\centering\includegraphics[width=0.8\linewidth]{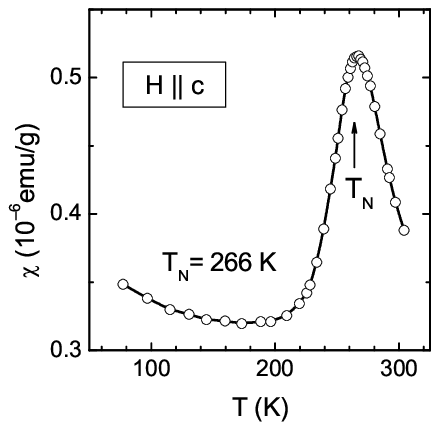}
\centerline{Figure 1 to paper Belevtsev et al.}
\end{figure}

\begin{figure}
\centering\includegraphics[width=0.8\linewidth]{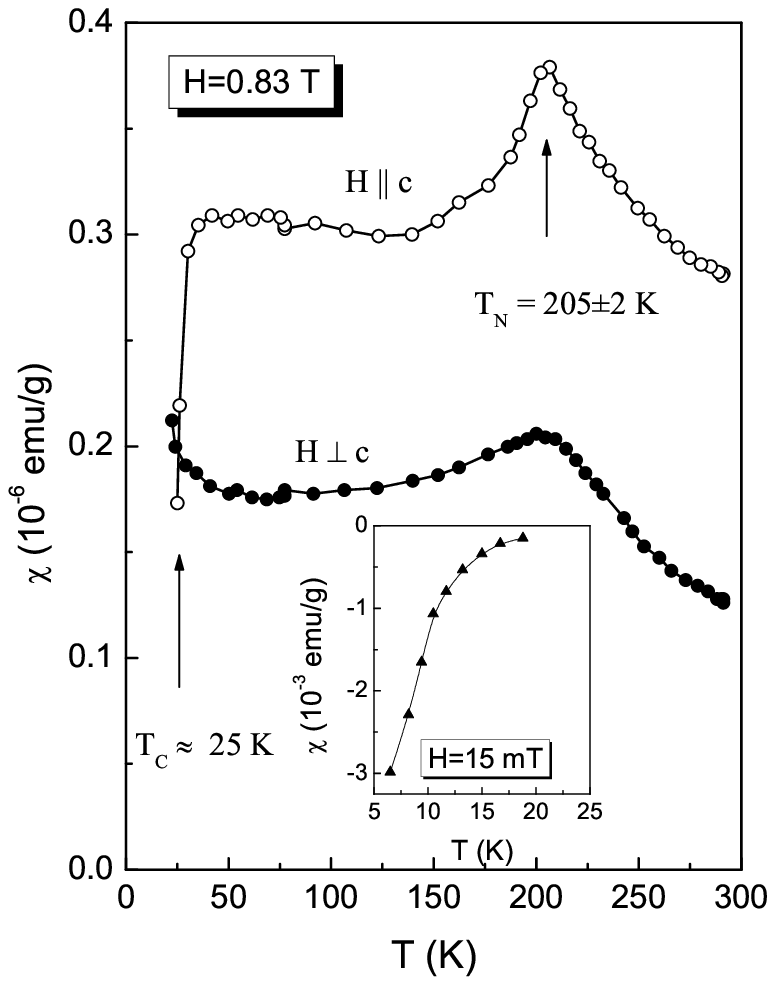}
\centerline{Figure 2 to paper Belevtsev et al.}
\end{figure}

\begin{figure}
\centering\includegraphics[width=0.8\linewidth]{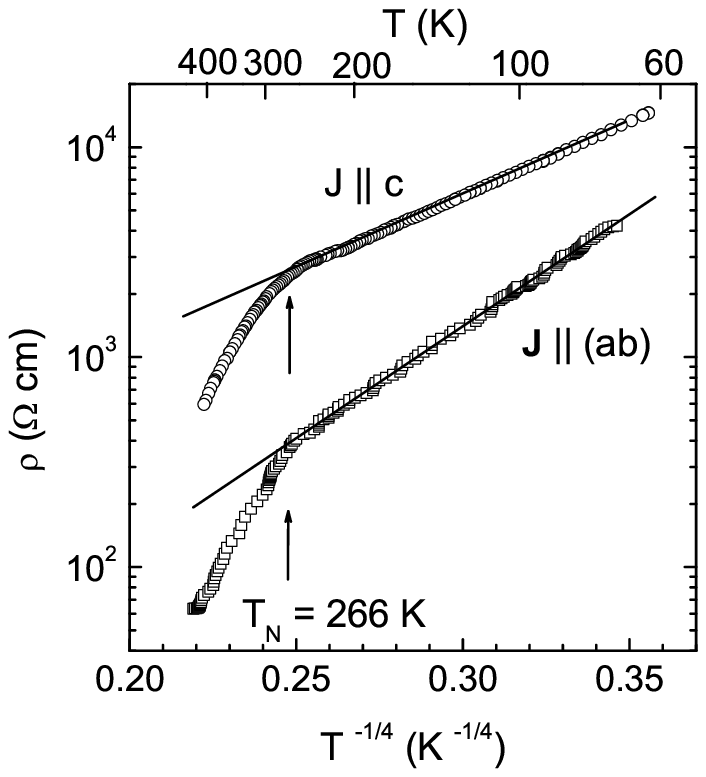}
\centerline{Figure 3 to paper Belevtsev et al.}
\end{figure}

\begin{figure}
\centering\includegraphics[width=0.8\linewidth]{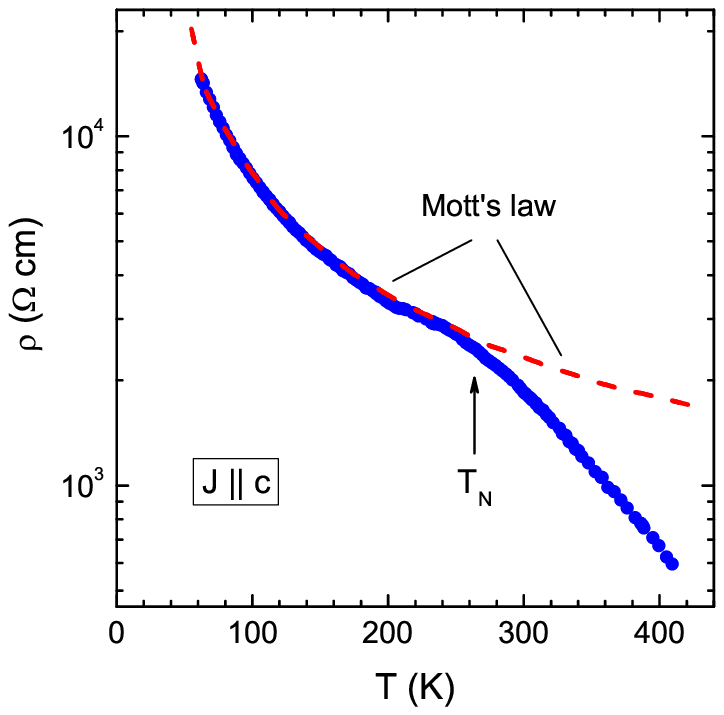}
\centerline{Figure 4 to paper Belevtsev et al.}
\end{figure}

\begin{figure}
\centering\includegraphics[width=0.8\linewidth]{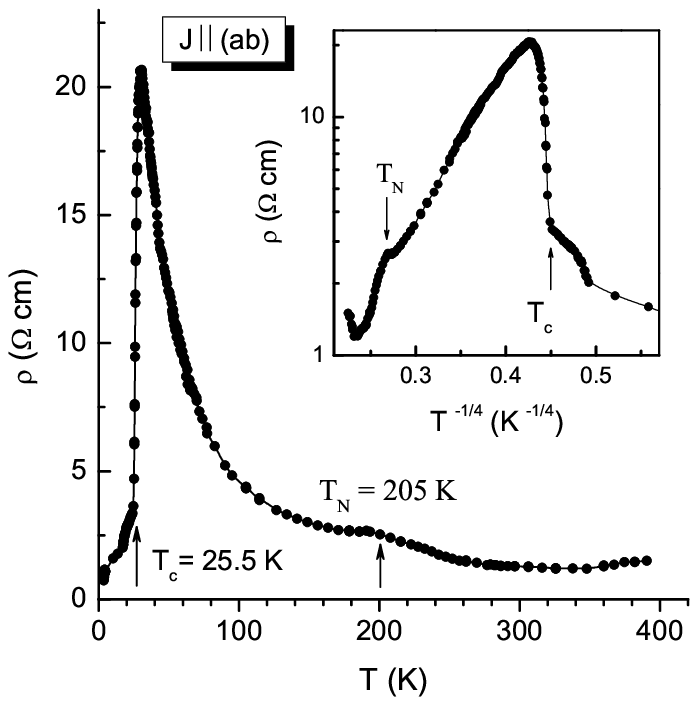}
\centerline{Figure 5 to paper Belevtsev et al.}
\end{figure}

\begin{figure}
\centering\includegraphics[width=0.8\linewidth]{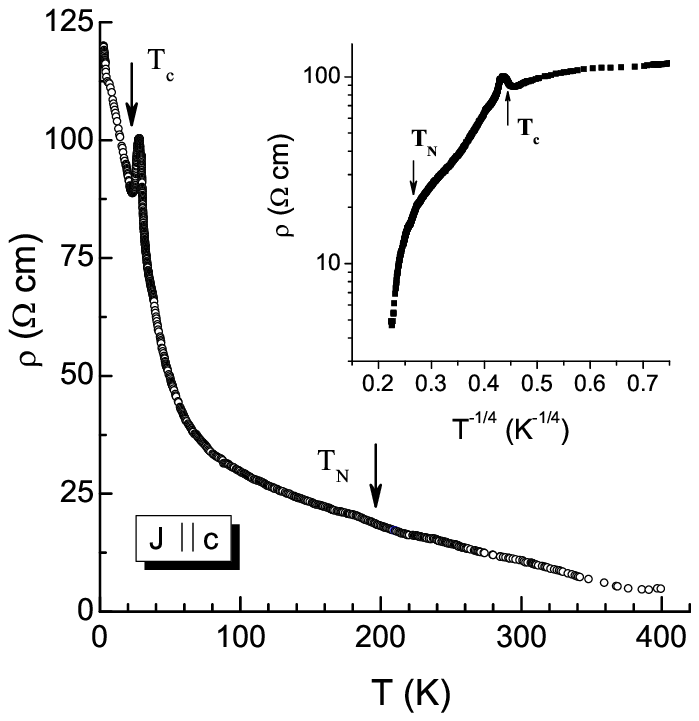}
\centerline{Figure 6 to paper Belevtsev et al.}
\end{figure}

\end{document}